\documentstyle[12pt]{article}
\def\go{
\mathrel{\raise.3ex\hbox{$>$}\mkern-14mu\lower0.6ex\hbox{$\sim$}}
}
\def\lo{
\mathrel{\raise.3ex\hbox{$<$}\mkern-14mu\lower0.6ex\hbox{$\sim$}}
}
\begin{document}
\title{ACCELERATION OF ULTRA HIGH ENERGY COSMIC RAYS}
\author{R. D. Blandford\\
130-33 Caltech\\
Pasadena CA 91125 USA} 
\maketitle
\begin{abstract}
Some general features of cosmic ray acceleration are summarized along
with some inferences that can be drawn concerning the origin of the UHE component.
The UHE luminosity density is found to be similar to that derived
for GeV cosmic rays and its slope suggests a distinct 
origin. Reports of clustering on small angular scale, if confirmed, 
would rule out most proposed source models.
More generally, it is argued that the highest energy particles can only be accelerated
in sites that can induce an EMF ${\cal E}\go3\times10^{20}$~V 
and an associated power 
$L_{{\rm min}}\go{\cal E}^2/Z\sim10^{39}$~W, where $Z$
is the characteristic, electrical impedance, typically $\lo100\Omega$. 
Shock acceleration, unipolar induction and magnetic flares are the 
three most potent, observed, acceleration mechanisms and radio jet termination shocks, 
$\gamma$-ray blast waves, dormant black holes in galactic
nuclei and magnetars are the least implausible, ``conventional''
manifestations of these mechanisms that have been invoked to explain the 
UHE cosmic rays. Each of these models presents problems and 
deciding between these and ``exotic'' origins for UHE cosmic rays,
including those involving new particles or defects will require improved
statistical information on the energies, arrival times and directions,
as should be provided by the Auger project. 
\end{abstract}
\section{INTRODUCTION}
I have been asked to summarize ``conventional'' schemes for the acceleration
of UHE cosmic rays, though any physical process capable of endowing
a subatomic particle with the kinetic energy of a 
well-hit baseball/cricketball can hardly be considered
conventional.  This means that I shall leave others to review mechanisms that attribute the 
origin of these particles to topological defects, strings, monopole decay, supersymmetric hadrons, 
cosmic necklaces, cryptons and so on. Indeed, I suspect 
that the ``hidden agenda'' is for me to fail at my appointed
task, and, like my colleagues on the MACHO experiment, to make the world safe for elementary
particle theorists. I shall not disappoint.

Many of the issues that I will cover have been recognised for some time and have 
been well-discussed in several excellent reviews including Hillas (1984)
and Cronin (1996) and the many relevant contributions to the recent
conference on this subject (Krizmanic, Ormes \& Streitmatter 1998),
including, especially, the lively summary by the late David Schramm.
The conference proceedings edited by Chupp \& Benz (1994) is also relevant.
\section{THE COSMIC RAY SPECTRUM}
In order to give this topic some context, consider the complete cosmic ray 
spectrum ({\it eg} Berezinski {\it et al} 1990).  
This extends over nearly twelve decades of energy 
from the proton rest mass, $\sim1$~GeV, where their energy density is 
that of the microwave background, to at least $300$~EeV
($\equiv50$~J $\equiv3\times10^{-8}$~m$_{{\rm Pl}}$). We can consider the 
cosmic ray spectral energy density inferred at the solar system, 
$U(E)=(4\pi/c)(dI/d\ln E)\sim5\times10^{-14}(E/{\rm 10GeV})^{-0.7}$~J m$^{-3}$
(correcting for solar modulation) extending from $\sim10$~GeV
to the ``knee'' at $\sim100$~TeV- 10 PeV. (10 GeV cosmic rays 
are about 10~m apart and have an energy density comparable
with that of the microwave background. The spectrum steepens above the knee:
$U(E)\sim4\times10^{-18}(E/10{\rm PeV})^{-1.1}$~J m$^{-3}$. It
then dips and flattens around the ``ankle'' ($\sim1-10$~EeV).
UHE cosmic rays - the toenail clippings of the universe - 
are observed up to 300~EeV and, with a little imagination, 
$U(E)\sim1.5\times10^{-21}(E/10{\rm EeV})^{-0.5}$~J m$^{-3}$,
comparable with the estimated, integrated background from $\gamma$-ray bursts. 
(Despite the large uncertainty, and the fact that the number density has
fallen by $\sim10$ orders of magnitude, we do measure the EeV spectrum 
better than the MeV spectrum, of which we are, quite decently, ignorant.)

The $\sim1$~GeV-$100$~TeV cosmic rays are of Galactic origin. 
The ratio of Li, Be, B secondaries to C, N, O primaries measures their range to be
$\lambda(E)\sim100(E/10{\rm GeV})^{-0.6}$kg m$^{-2}$ ({\it eg} Axford 1994). 
The cosmic ray luminosity of the Galaxy is then estimated as 
$\sim M_dU(E)c/\lambda(E)\sim2\times10^{33}(E/1{\rm GeV})^{-0.1}$~W 
where $M_d$ is the gas mass of the disk. Scaling from the local 
galaxy luminosity density (per $\ln E$ and assuming $h=0.6$), 
we derive an average, cosmological,  luminosity density (per $\ln E$),
$<{\cal L}>(E)\sim4\times10^{-37}(E/10{\rm Gev})^{-0.1}$~W m$^{-3}$, 
for $10{\rm GeV}<E<100{\rm TeV}$. (For comparison,
the stellar luminosity density is $\sim10^{-33}$~W m$^{-3}$.)

The UHE particles are almost surely extragalactic. As with $\gamma$-ray bursts, 
there is no good evidence for disk, 
halo, cluster or supercluster anisotropy (despite some tantalising hints in the past),
(Takeda {\it et al} 1999). Furthermore, magnetic confinement by the Galaxy is impossible - 
the Larmor radius $r_L(E)$ of a 300~EeV cosmic ray in a $\mu$G field is 
$\sim300$kpc. If we assume that UHE cosmic rays are protons, 
(and assuming that they are not, only makes 
matters worse), then they have a short lifetime to photo-pion 
production on the microwave background, 
(Greisen 1966, Zatsepin \& Kuzmin 1966). The characteristic lifetime of a 
$\sim60-300$~EeV cosmic ray is, very roughly, 
$T(E)\sim0.1(E/300{\rm EeV})^{-2}$~Gyr. This implies that the luminosity 
density increases with energy 
$<{\cal L}>(E)\sim U(E)/T(E)\sim10^{-37}(E/300{\rm EeV})^{1.5}$~W m$^{-3}$.
At the highest measured energy, the estimated cosmological luminosity density is not significantly 
different from that of  10 GeV cosmic rays.
The change in slope in the source spectrum, above $\sim60$~EeV, is a 
strong indication that these UHE cosmic rays comprise a 
quite distinct component from their lower energy counterparts.

In order to investigate this further, it is necessary to take account of the
fluctuations in energy loss. Taking the 17 
events reported by the AGASA collaboration above 60~EeV, it is possible to derive 
a maximum likelihood estimate of the unnormalized
energy density, uncorrected for biases in detection efficiency.
I find that if $U(E)\propto E^\alpha;E>E_{{\rm min}}=60$~EeV, then $\alpha=-1.2\pm0.5$.
I then calculate the probability that a particle of energy $E_0$
has energy $>E$ after time $t$, $P(E,t;E_0)$, following Aharonian \& Cronin (1994) 
and Bahcall \& Waxman (1999). If we assume
a power law for the luminosity density ${\cal L}(E)\propto E^\beta;E>E_{{\rm min}}=60$~EeV,
then the logarithm of the likelihood for obtaining the observed events is 
\begin{equation}
\propto\sum_i\ln\left[(1-\beta)E_{{\rm min}}^{(1-\beta)}\int_{E_i}dE_0E_0^{\beta-2}
\int dt{\partial P\over\partial\ln E}\right]
\end{equation}
Maximizing this function with respect to variation of 
$\beta$, gives the estimate $\beta=0.3\pm0.2$. A more sophisticated
computation that takes into account the detection probabilities 
of the different events should be performed, 
but it is unlikely to change the conclusion
that the spectral luminosity density actually increases in the 
60-300 EeV energy range and may even be consistent with a single, ``top down''
source with energy well above 300 EeV.

There have been reports that UHE cosmic rays are significantly clustered on the sky.
Specifically, in a sample of 47 events observed with AGASA, there are three pairs
and one triple above 40 EeV with separations $\lo2.5^\circ$,
comparable with the positional errors (Takeda {\it et al} 1999).  
(There are two more coincidences with events drawn from other samples.) 
There is no clear pattern for the associated particles 
to be ordered in energy and, in particular, one double has a 106EeV
particle arriving over 3 yr. after a 44 EeV particle.  

{\it If} these associations are real, then there are three 
important implications. Firstly, as particles are likely to be deflected by intergalactic 
magnetic field through an angle $\delta\theta\sim(D\ell_B)^{1/2}/r_L(E)$
then they will be delayed by $\sim D^2\ell_B/r_L(E)^2c\propto E^{-2}$, where $\ell_B$ is the field
correlation length and $D\sim cT(150{\rm EeV})\sim30$~Mpc is the supposed source distance
({\it} cf Miralda-Escud\'e \& Waxman 1996). Even Aesop would be challenged to 
explain how a $\sim40$~EeV cosmic ray precedes a $\sim100$~EeV cosmic ray
if they started at the same time 
and we must conclude that the source persists for several years, at least. This
would rule out all particle/defect and $\gamma$-ray burst models. Secondly,
the small deflection angles at low energy limit the intergalactic 
field strength to $B\lo20(\ell_B/1{\rm Mpc})^{-1/2}$~fT, far 
smaller than generally supposed, though probably not excludable by 
direct observation. Thirdly, the presence of three $\sim40$~EeV cosmic
rays associated with high energy cosmic rays of much shorter range, implies that the background 
of low energy cosmic rays {\it not} associated with high energy events must be larger 
than the incidence of clustered events by roughly the
ratio of their typical lifetimes $\sim30$, which more than accounts for the remainder of the 
low energy sample.
This, in turn, implies that the high energy cosmic rays must come from a very few sources which are, 
consequently, quite energetic: $E\sim10^{44}(\tau/3{\rm yr})$~J, where $\tau$ is their lifetime.
(If the low energy cosmic rays are scattered through $\sim2.5^\circ$, then $\tau\go10^5$~yr
and $E\sim3\times10^{48}$~J.)

However, this clustering hypothesis, which is necessarily {\it a posteriori}
when expressed in detail, is only supported with modest confidence. (A simple, Monte Carlo
simulation distributing 47 points at random on half the sky and looking for similar
patterns is quite instructive.) Particle/defect/burst 
explanations of UHE cosmic rays need not yet be rejected on these grounds.
\section{COSMIC RAY ACCELERATION}
The standard model of bulk cosmic ray production is first order 
Fermi acceleration at strong, super-Alfv\'enic, shocks associated 
with supernova remnants and, possibly,
winds from hot stars ({\it eg} Blandford \& Eichler 1987).  
A typical relativistic proton will cross a shock, travelling
with speed $u$, $O(c/u)$ times, gaining energy $\Delta E/E=O(u/c)$ each traversal through
scattering by hydromagnetic waves moving slowly with respect to the converging fluid flows on either
side of the front.  The net mean relative energy gain is $O(1)$, but the process is statistical
and a kinetic calculation shows that the transmitted spectrum will be a power
law in momentum, $f(p)\propto p^{-3r/(r-1)}$, where $r$, ($=4$ for a strong shock), is the compression
ratio. This mechanism can account, broadly, for the power (eg Malkov 1999), the 
slope ({\it eg} Axford 1994) and the composition ({\it eg} Ellison {\it et al} (1997))
of GeV cosmic rays. Shock acceleration is also,
arguably, observed directly in SN1006 (Koyama {\it et al} 1995), as well as in the solar system 
({\it eg} Erd\"os \& Balogh 1994).

The maximum energy to which a particle can be accelerated at a shock front is dictated by the scattering
mean free path, $\ell(E)\sim(B/\delta B)_{r_L}^2r_L(E)$, where $\delta B$ is the amplitude of
resonant hydromagnetic waves with wavelength matched to the particle Larmor radius. 
The diffusion scale-length
of cosmic rays ahead of the shock is $\sim\ell c/u$ and, assuming that this is limited
by the size of the shock $\sim R$ we arrive at the unsurprising result that the maximum
energy achievable in shock acceleration, assuming $\delta B<B$ and the presence of
a large scale magnetic field, is $E_{{\rm max}}=e{\cal E}\sim euBR\sim ed\Phi/dt$, the product of the 
charge and the motional potential difference across the whole shock. Equivalently, we conclude
that in order to accelerate a proton by this mechanism to an energy ${\cal E}e$, the rate
of dissipation of energy exceeds $L_{{\rm min}}\sim {\cal E}^2/Z$, where 
$Z=\mu_0u=\mu_0E/B$ is the effective impedance of the accelerator in SI units.

Imposing this condition for a supernova remnant in the interstellar medium
leads to an estimate $E_{{\rm max snr}}\sim30$~TeV ({\it eg} Axford 1994), close to the knee. 
An additional source is needed between the knee and the ankle, where the source
is generally supposed to be metagalactic. Larger shocks, especially those at Galactic wind
termination shocks ({\it eg} Jokipii \& Morfill 1987)
and associated with gas flows around groups and clusters
of galaxies have been invoked ({\it eg} Norman {\it et al} 1995). 
These shocks are likely to be relatively weak and 
therefore to transmit steeper spectra, as observed.  The major uncertainty is the strength of 
the magnetic field. If $B\sim30$~pT at a galactic shock and $\sim10$~pT at a 
cluster shock, then $E_{{max}}\sim10,100$~PeV respectively. Neither site is likely to 
accelerate the highest energy particles.

An alternative accelerator is the unipolar inductor ({\it eg} Goldreich \& Julian 
1967). The archetypical example is a pulsar
- a spinning, magnetised, neutron star. The surface field will be quite complex
but a certain quantity of magnetic flux $\Phi$ can be regarded as ``open''
and tracable to large distances from the star, (well beyond the light cylinder).
As the star is an excellent conductor, an EMF
will be electromagnetically induced across these open field lines ${\cal E}\sim\Omega\Phi$,
where $\Phi$ is the total, open magnetic flux. This EMF will cause currents to flow 
along the field and as the inertia of the plasma is likely to be insignificant
the only appreciable impedance in the circuit is related to 
the electromagnetic impedance of free
space $Z\sim0.3\mu_0c\sim100$~$\Omega$.  The maximum energy to which a particle can be 
accelerated is $E_{{\rm max}}\sim e{\cal E}$ and the total rate at which energy is extracted
from the spin of the pulsar is $L_{{\rm min}}\sim{\cal E}^2/Z$.  Taking the Crab pulsar
as an example, $E_{{\rm max}}\sim30$~PeV for protons
and $L_{{\rm min}}\sim10^{31}$~W. As the stellar surface may well comprise iron, even the Crab pulsar
has the capacity to accelerate up to EeV cosmic rays. However, it is not obvious that 
all of this potential difference will actually be made available for particle acceleration. 
In particular, this is unlikely to happen in the pulsar magnetosphere
as a large electric field parallel to the magnetic field will be shorted out
by electron-positron pairs, which are very easy to produce, and radiative drag is likely
to be severe. A more reasonable site is
the electromagnetic pulsar wind and the surrounding nebula where particles can gain energy as they
undergo gradient drift between the pole and the equator (Bell 1992). Pulsars may well 
contribute to the spectrum of intermediate energy cosmic rays.

A third, protoypical accelerator is a flare, for example one occuring on the solar surface
or the Earth's magnetotail.
Here magnetic instability leads to a catastrophic rearrangement, which must be 
accompanied by a large inductive EMF. Unless the instabilities
are explosive, the effective impedance is again $\sim\mu_0 u$, where $u$ is a characteristic 
speed. Non-relativistic flares generally convert most of the dissipated magnetic energy into 
heat and are notoriously inefficient in accelerating high energy particles.

Other acceleration mechanisms have been proposed and may contribute to the acceleration of the
bulk of Galactic cosmic rays and relativistic electrons in
non-thermal sources. These include a variety of second order
processes and steady, magnetic reconnection. Many of them can be observed to operate within
the solar system.  However, they are thought to be too slow to be relevant to the
acceleration of the highest energy cosmic rays. 
\section{ZEVATRONS}
Having argued that the three most potent, observed acclerators are shocks,
unipolar inductors and flares, let us see how they can be modified to account for
$\sim$~ZeV cosmic rays.  Firstly, note that, as $u\sim c$, mildly relativistic shocks minimise 
the power that has to be invoked to attain high energy. Specifically,
we need a power $>10^{39}$~W to account for 300 EeV cosmic rays and this exceeds
the bolometric luminosity of a powerful quasar. One of the few sites 
where such a large potential difference can be achieved is the termination
shock of a powerful radio jet like that associated with Cygnus A ({\it eg} Cavallo
1978).  Stretching the numbers
a little, we combine a field strength $\sim10$~nT, with a speed $\sim c$ and a transverse scale
$\sim3$~kpc which gives $E_{{\rm max}}\sim300$~EeV. The problem with this model is that 
observed UHE cosmic rays are not positionally identified with the few known radio sources
within $D(E)\sim30$~Mpc that might be powerful enough to account for them 
({\it cf} Farrar \& Biermann 1998).  

A more elaborate shock accelerator is the $\gamma$-ray burst blast wave
({\it eg} Waxman 1995).
Here, the shocks (assumed to be spherical) are ultrarelativistic with Lorentz factor $\Gamma$. 
The maximum energy, measured in the frame of the explosion,
to which a proton can be accelerated in a dynamical timescale from an 
ultrarelativistic shock of radius $R$ is 
$E_{{\rm max}}\sim eB'Rc$, where $B'$ is the comoving field strength.  The explosion power,
adopting the most elementary of assumptions, is then $L_{{\rm min}}\sim4\pi\Gamma^2
(E_{{\rm max}}/e)^2/\mu_0c$. Observed bursts have typical explosion powers estimated
to be $L_{{\rm exp}}\sim10^{45}$~W, which can be consistent  with 300~EeV proton acceleration
as long as $\Gamma<300$, which is just compatible with 
existing models. A serious physical constraint is the avoidance of radiative loss in this
environment. An observational concern with this model is the improbability of having enough active
bursts close to supply the highest energy particles
roughly isotropically ({\it cf} Waxman \& Miralda-Escud\'e 1996).

The most relevant variant on unipolar induction is 
magnetic energy extraction from spinning, black holes, where 
the magnetic field is supported by external current, and the horizon 
is an imperfect conductor with resistance $\sim100\Omega$ (eg Thorne
{it et al} 1986). This impedance
is matched to the electromagnetic load so that roughly
half of the available spin energy ends up in the irreducible 
mass of the hole, the remainder being made available for particle acceleration.
The total electromagnetic power needed to account for $\sim300$~EeV acceleration
is, once more, $\sim10^{39}$~W. A rapidly spinning, 
$\sim10^9$~M$_\odot$ hole endowed with a field strength $\go1$~T or a 
$\sim10^5$~M$_\odot$ hole threaded by a $\go10^4$~T field suffices to
accelerate 300EeV particles. The major concern with this model is 
that the radiation background must be extremely low in order that catastrophic loss
due to pion and pair production be avoided.  Specifically,
it is necessary that the microwave luminosity in an acceleration zone, of size $R$,
be $\lo10^{34}(R/10^{14}{\rm m})$~W, far smaller than the 
unobserved electromagnetic power.

The best generalization of flare acceleration involves ``magnetars''
which are young, spinning neutron stars
endowed with a $\sim10-100$GT surface magnetic field as first postulated by
Thompson \& Duncan, (1996). 
Now, the observation of $5-7$~s period pulsations from three ``soft gamma repeaters''
effectively confirms their identification as old magnetars that have been decelerated 
by electromagnetic torque and which are now powered by magnetic energy which is 
released in a series of giant flares, Kouveliotou {\it et al} 1998.
The inductive EMFs associated with an electromagnetic flare from a magnetar
can be as high as
${\cal E}\sim3\times10^{19}$~V, making them candidate UHE accelerators 
because the surface composition is likely to be Fe.  However, the available reservoir of 
magnetic energy is only $\sim10^{40}$~J and the magnetar birthrate is 
no more than $\sim10^{-3}$~yr$^{-1}$ in the Galaxy. This rules them out as
an extragalactic source.  Only if UHE cosmic rays have a Galactic origin, (and the 
large scale anisotropy observations suggest quite strongly that do not),
can there be enough power in magnetars to account for the UHE energy density. 
\section{DISCUSSION}
I have argued, tentatively, that UHE cosmic rays are created 
in a new population of extragalactic sources with an average luminosity density that approaches
that of Galactic cosmic rays. I have also described problems 
with each of the candidate ``conventional'' mechanisms for 
accelerating protons to these high energies.
Quite different, general inferences have been drawn here,
from the same data, by Waxman, and elsewhere by others. All of this 
underscores the need for better statistics which should be met by the Auger
project. Perhaps the most pressing need 
is to understand if particles of very different energy have a common origin. 
If true, this must rule 
out essentially all primordial particle/topological defect, neutron star, $\gamma$-ray
burst explanations, leaving only massive black holes and radio source 
models among the possibilities discussed above. In this case, it will be possible
to seek identifications, 
especially at the highest energies, where the positions will be most accurate and the 
delays due to magnetic scattering the smallest. If, alternatively, 
clustering and its implications are {\it not} substantiated,
then the next best clues will probably come from composition studies and detailing
the large scale distribution on the sky. 

The most exciting outcome of all of this is that we are dealing with a new particle
or defect with energy well out of the range of terrestrial accelerators. 
(For example, if there is a particle of energy $E_X$ which decays 
with half life $\tau_X$ into $N$ protons, then the cosmological energy density
of these particles must be $\Omega_X\sim3\times10^{-8}N^{-1}
(E_X/1{\rm YeV})^{-1}(H_0\tau_X)^{-1}$.)
Whatever happens, in a subject where the
dullest and most conventional theories involve massive, spinning, black holes,
ultrarelativistic blast waves and 100 GT fields threading nuclear matter, 
the future is guaranteed to be interesting. 
\section*{Acknowledgements}
I am indebted to John Bahcall, Jim Cronin, Michael Hillas, Martin Rees, Alan Watson and Eli Waxman
for stimulating discussions and the editors for their forbearance. I also gratefuly acknowledge
the hospitality of the Insitute for Advanced Study (through the Sloan
Foundation) and the Institute of Astronomy (through the Beverly and Raymond Sackler Foundation)
as well as NASA grant 5-2837.

\end{document}